\documentclass{evn2024}
\usepackage{graphicx}
\usepackage{url}
\usepackage{natbib}
\begin{document}
   \title{The GRACE project}
   \subtitle{Hard X-ray giant radio galaxies and their duty cycle}
   \author{G. Bruni\inst{1}, F. Panessa\inst{1}, L. Bassani\inst{2},  M. Brienza\inst{3}, M. Fanelli\inst{1,4}, F. Ursini\inst{5}, F. Massaro\inst{6,7}, A. Malizia\inst{2},\\ M. Molina\inst{8}, L. Hern\'andez-Garc\'ia\inst{9,10,11}, C. J. Riseley\inst{12,3}, E. K. Mahony\inst{13}, M. Janssen\inst{14,15},\\ D. Dallacasa\inst{12,3}, T. Venturi\inst{3}, R.~D. Baldi\inst{3} \and M. Persic\inst{16} 
    }

   \institute{
        INAF -- Istituto di Astrofisica e Planetologia Spaziali, via del Fosso del Cavaliere 100, Roma, 00133, Italy
        \and
        INAF -- Osservatorio di Astrofisica e Scienza dello Spazio di Bologna, via Piero Gobetti 93/3, 40129 Bologna, Italy
        \and
        INAF -- Istituto di Radioastronomia, Via Gobetti 101, 40129 Bologna, Italy
        \and
        Dipartimento di Fisica, Sapienza Università di Roma, P.le A. Moro 5, Roma 00185, Italy
        \and
        Dipartimento di Matematica e Fisica, Università degli Studi Roma Tre, via della Vasca Navale 84, 00146 Roma, Italy
        \and
        Dipartimento di Fisica, Università degli Studi di Torino, Via Pietro Giuria 1, 10125 Torino, Italy
        \and
        Istituto Nazionale di Fisica Nucleare, Sezione di Torino, Via Pietro Giuria 1, 10125 Torino, Italy
        \and
        INAF -- Istituto di Astrofisica Spaziale e Fisica cosmica, via Alfonso Corti 12, 20133 Milano, Italy
        \and
        Millennium Nucleus on Transversal Research and Technology to Explore Supermassive Black Holes (TITANS) 
        \and
        Millennium Institute of Astrophysics (MAS), Nuncio Monseñor Sótero Sanz 100, Providencia, Santiago, Chile
        \and 
        Instituto de F\'isica y Astronom\'ia, Facultad de Ciencias,Universidad de Valpara\'iso, Gran Bretana 1111, Playa Ancha, Valpara\'iso, Chile
        \and
        Dipartimento di Fisica e Astronomia, Università degli Studi di Bologna, Via P. Gobetti 93/2, 40129 Bologna, Italy 
        \and
        Australia Telescope National Facility, CSIRO Space and Astronomy, PO Box 76, Epping, NSW 1710, Australia
        \and
        Department of Astrophysics, Institute for Mathematics, Astrophysics and Particle Physics (IMAPP), Radboud University, P.O. Box 9010, 6500 GL Nijmegen, The Netherlands
        \and
        Max-Planck-Institut f\"ur Radioastronomie, Auf dem H\"ugel 69, D-53121 Bonn, Germany
        \and
        INAF -- Osservatorio Astronomico di Padova, vicolo dell'Osservatorio 5, 35122 Padova, Italy
             }

   \abstract{
The advent of new generation radio telescopes is opening new possibilities on the classification and
study of extragalactic high-energy sources, specially the underrepresented ones like radio galaxies.
Among these, Giant Radio Galaxies (GRG, larger than 0.7 Mpc) are among the most extreme
manifestations of the accretion/ejection processes on supermassive black holes. Our recent
studies have shown that GRG can be up to four times more abundant in hard X-ray selected
(i.e. from INTEGRAL/IBIS and Swift/BAT at $>$20 keV) samples and, most interestingly, the
majority of them present signs of restarted radio activity. This makes them the ideal test-bed to
study the so far unknown duty cycle of jets in active galactic nuclei. Open questions in the field
include: How and when jets are restarted? How jets evolve and what’s their dynamic? What is
the jet’s duty cycle and what triggers them? Our group has recently collected a wealth of radio
data on these high-energy selected GRGs, allowing us to study their jet formation and evolution
from the pc to kpc scales, across different activity epochs. In particular, thanks to our EVN large
programme, we were able to probe the new radio phase in the core of these giants. Furthermore,
we are devoting an effort to the exploitation of new radio surveys data for the discovery of new
classes of counterparts of Fermi/LAT catalogues. In particular, we are unveiling
the hidden population of radio galaxies associated with gamma-ray sources.
   }
   \titlerunning{The GRACE project}
   \authorrunning{G. Bruni et al.}
   \maketitle
%

\section{Introduction}
Giant Radio Galaxies (GRG, \citealt{Willis+1974}) are among the most extreme manifestations
of the accretion/ejection processes on supermassive black holes. During their $\sim$100
Myr time-scale evolution \citep{machalski2004,Jamrozy+08}, they produce jets of plasma extending hundreds
of kpc away from their active galactic nucleus (AGN) core. Their projected linear
size, known to be between a conventional threshold of 0.7 Mpc and a maximum of 4-5 Mpc
for decades (3C236, J1420-0545, and Alcyoneus, see \citealt{Schoenmakers2000,Machalski+08,2022A&A...660A...2O}), was recently extended to 7 Mpc (Porphyrion, \citealt{2024Natur.633..537O}). This implies that their jets are not only able to probe the intergalactic medium, but also fill voids between galaxy clusters. With such a long activity period, GRGs are
the ideal testbed to study the duration of the radio phase in AGN, and its duty cycle.
The recent release of new generation radio surveys (e.g. LoTSS, VLASS,
RACS) is opening new possibilities for the study of radio galaxies, allowing
for a deeper and sharper search, increasing their census even in the high energy domain, so far dominated by blazars.

Since early 2000s, the International Gamma-Ray Astrophysics Laboratory (INTEGRAL) and the Neil Gehrels Swift Observatory (Swift) have been scanning the hard X-ray sky (20$\sim$100 keV) at a sensitivity better than a few mCrab, and a point source localization accuracy of a few arcmin. The missions have produced catalogues of hard X-ray selected Active Galactic Nuclei (AGN, e.g. \citealt{Malizia_update}, \citealt{swift100}, and references therein) offering an unbiased view on the high-accretion regime on supermassive black holes. The samples have been the subject of several studies along years \citep{malizia2014,panessa13,panessa2015,panessa2016}. As expected, radio galaxies are only a fraction of these samples ($\sim$8\%). Despite their rarity, they offer the unique possibility to study at the same time ejection and accretion processes, and their connections.


\begin{figure*}
    \centering
    \includegraphics[width=0.9\linewidth]{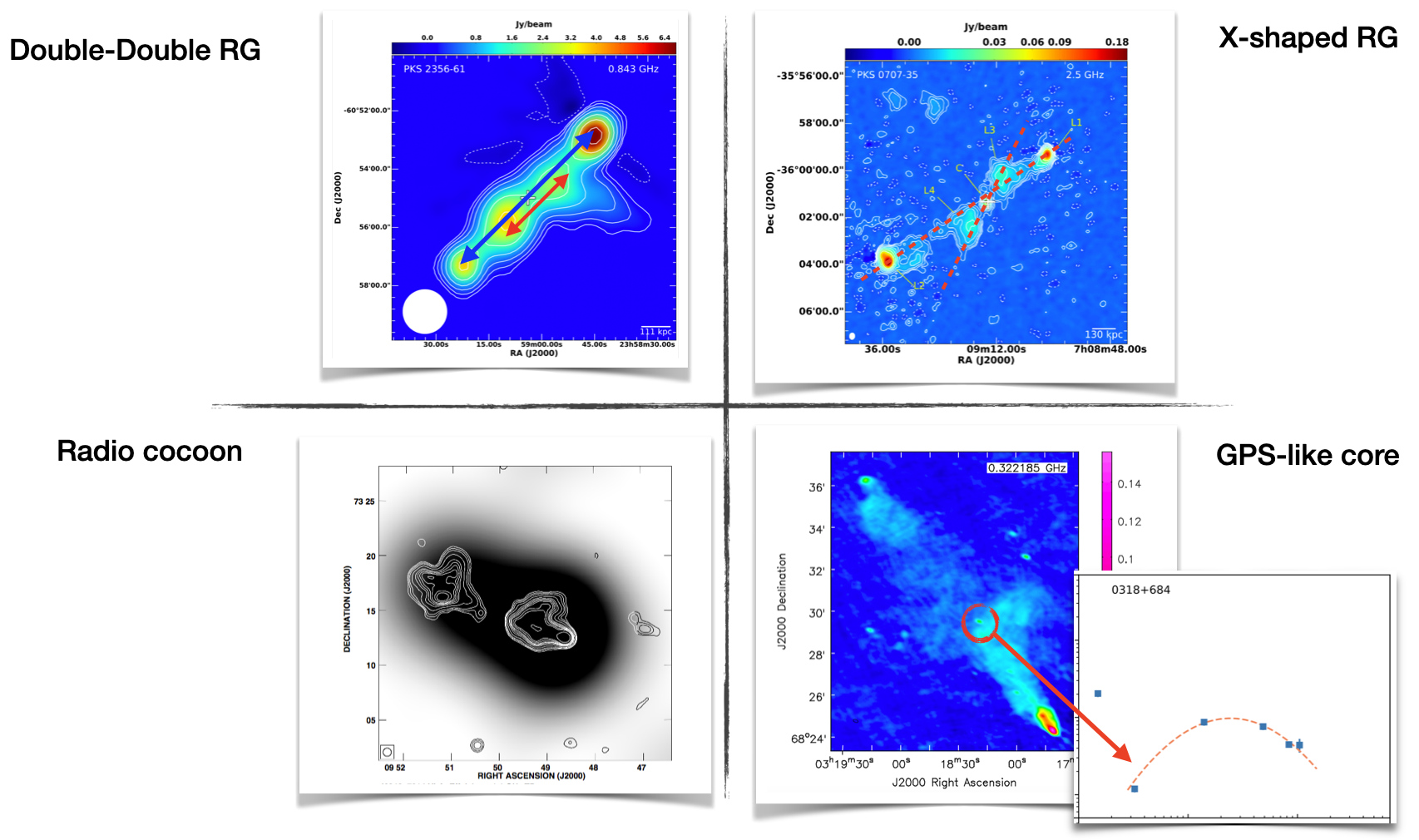}
    \caption{Main indicators of restarted radio activity: a double-double and an X-shaped moprhology, where two couples of lobes are present, even on different axes, a radio cocoon, fossil of a previous radio phase, a GPS-like core, showing a peaked spectrum.}
    \label{restarting}
\end{figure*}


\section{The GRACE project}
Recently, our group (GRAL - Gamma Radio Astrophysical Lab, \url{http://gral.iaps.inaf.it/}) has
been focusing on the investigation of radio emission in such sample \citep{chiaraluce2020,panessa2022}, and in
particular in radio galaxies. The latter show a fraction of GRG almost twice the one
found in traditionally radio-selected samples \citep{Bassani+2016}. Our follow-up studies on these
hard X-ray GRG (HXGRG), through both a morphological and spectral analysis (see Fig. \ref{restarting}, revealed the presence of a restarted radio activity in the majority of objects \citep{Bruni+2019,Bruni+2020,Bruni+2021}. This makes the sample the ideal testbed to study the jet duty cycle and triggering in AGN, so far still an open question for both stellar mass and supermassive black holes physics.

\subsection{The jets duty cycle in AGN: open questions}
We triggered a dedicated campaign involving the main radio telescopes around
the globe (the GRACE project - Giant RAdio galaxies and their duty Cycle, \url{https://
sites.google.com/inaf.it/grace/home}) securing datasets to answer the following open
questions:
\begin{itemize}
\item \textbf{How and when jets are restarted?} Zooming into their cores by means of
the VLBI techniques thanks to the collected data from our EVN and LBA
large project (PI Bruni), and e-MERLIN pilot observations (PI Bruni),
we aim at 1) studying the recently restarted radio phase 2) spotting
any hint of jet precession, and 3) probing the possible presence of a
binary supermassive black hole system (B-SMBH). Observations include
a comparison sample hard X-ray \textit{quiet} GRG (HQGRG), that will allow us
to put the results in the context, understanding the selection effect and
properties introduced by the HX selection. As an added value, the
latter goal would be a preparatory study for the next generation
of gravitational waves antennas (E.T., LISA), that will cover the
low frequency domain dominated by B-SMBH.

\item \textbf{How jets evolve and what’s their dynamic?} Study the Mpc-scale lobes
morphology to recover the information on the evolution and dynamics
of these sources on the Mega-years time scale. We will make use of
the already collected data from our VLA and GMRT campaign on a pilot
sample of 2 HXGRG (PI Bruni), plus LOFAR survey data available for
about half of the HXGRG and HQGRG comparison sample (LoTSS DR2,
recently released). This will allow us to individuate the different radio
phases, putting constraints on the duration of the radio activity along
the AGN activity. Our collaboration includes experts in MHD simulations
of jets, allowing for a comparison of the observed morphology with the
simulated ones: this will lead to an estimate of the physical parameters of
the plasma and environment.

\item \textbf{What is the jets duty cycle?} With the same dataset, it will be possible to
estimate the synchrotron ageing of the jets’ plasma through a modelling
of the radio SED for the different regions, allowing to date the plasma
ejection time and to recover information on the radio phases temporal
evolution.
\end{itemize}

As a whole, the study of the nature and evolution of the HXGRG and HQGRG
samples will help understanding how the hard X-ray selection of this sample could favour the discovery of merging, restarting, and in general multi-phase radio
galaxies, shedding light on the radio activity of AGN.


\begin{figure*}
    \centering
    \includegraphics[width=0.8\linewidth]{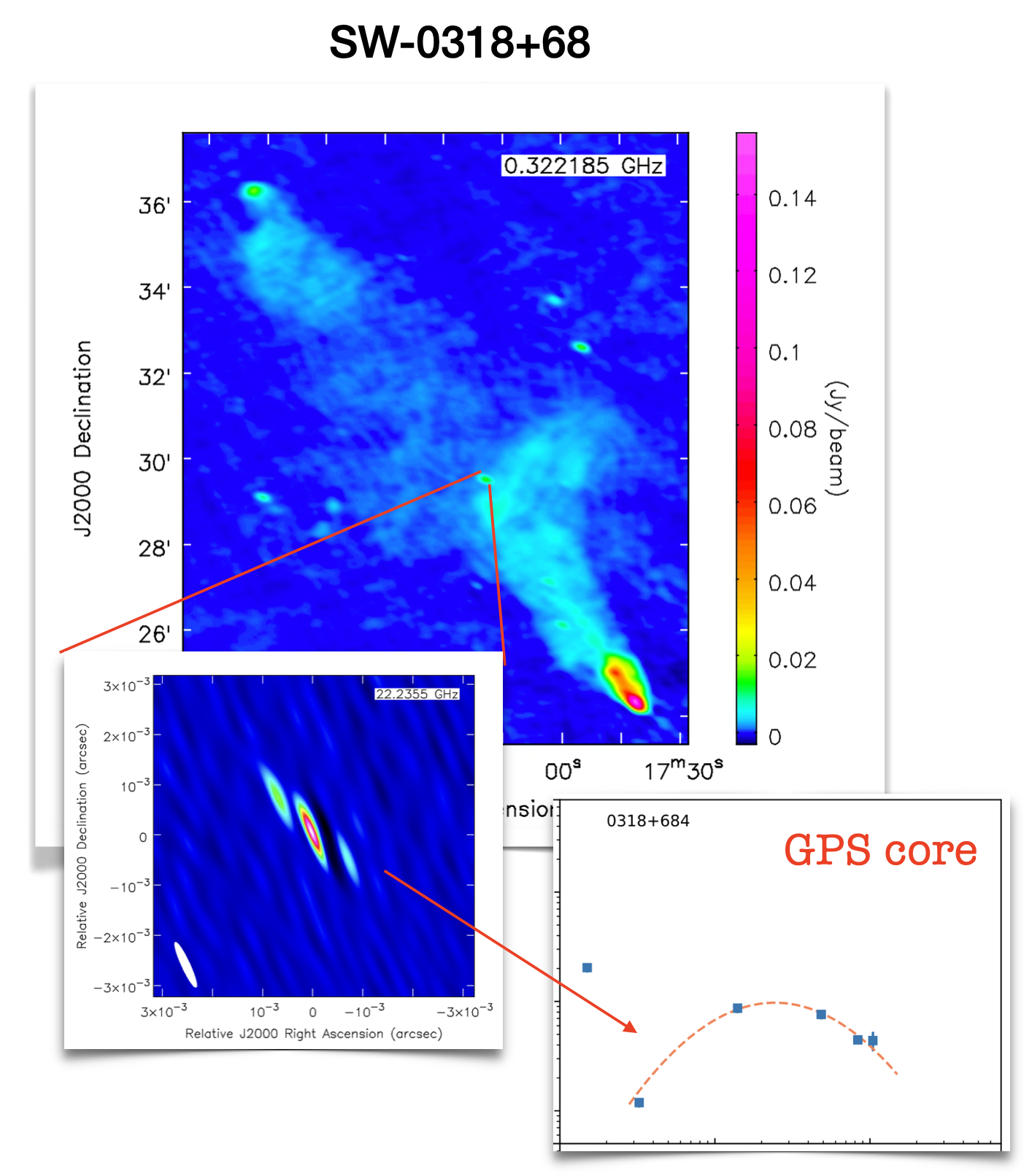}
    \caption{Multiscale imaging of the giant radio galaxy SW 0318+68. With a projected linear size of $\sim$1.6 Mpc, this is the most extended source of the sample. The Mpc-scale image is from the LoTSS survey (150 MHz, \citealt{Bruni+2021}), and shows the extended jets and lobes. The zoom into the pc-scale core region was realized with EVN observations at 8 GHz (contours) and 22 GHz (colors). Mini-lobes and the core are visible. The radio spectrum, showing a peak similar to GigaHertz-Peaked Sources (GPS), has been built with multi-frequency Effelsberg observations plus survey data \citep{Bruni+2019}.}
    \label{0318}
\end{figure*}


\subsection{First results from the VLBI campaign}
VLBI observations of the sample started with a peculiar source showing different classification in different bands, difficult to explain in the framework of the AGN unification scheme: PBC\,J2333.9-2343. Indeed, while on the Mpc scale it shows extended lobes, apparently on the plane of the sky, on the pc-scale a core-jet morphology was revealed by VLBA observations. A broad-band modeling could estimated a viewing angle of just a few degrees, and a multi-wavelength monitoring could confirm the strong variability typical of Blazars \citep{HernGarcia2017,2023MNRAS.525.2187H}. The fact that lobes are detached from their core, led to conclusion that the inner jet in the core region is most probably a new radio phase, where the jet axis dramatically changed its direction. The reason for its $\sim90^\circ$ reorientation is still unclear, although in such cases a merger is the main suspect.

These premises, together with the high fraction of restarted radio sources in the sample, suggested that the hard X-ray selection could more easily select extreme examples of reactivation or reorientation of the jet. Thus, in 2020 we carried out a combined EVN+LBA campaign to cover the entire sample at pc-scale resolution. In particular, observations at 8 and 22 GHz were designed to study the newborn radio phase in the cores showing a GigaHertz-Peaked Source (GPS) spectrum, and estimate the angle of the new jet with respect to the Mpc-scale lobes. First results suggest that all sources showing a GPS-like spectrum in their core also present mini-lobes on the pc-scale, confirming the restarted radio phase (see Fig. \ref{0318}). For other 2 sources (non-GPS) we found indications of a jet reorientation. 

Concluding, the co-presence of Mpc-scale lobes and a new radio phase (GPS) will allow us to estimate the duty cycle in these sources. First estimates via synchrotron aging of lobes suggest a value of $\sim$50-70 Myr, to be compared with the one of the GPS phase, usually a few thousands years \citep{ODea2021}.


\section{The tip of the iceberg: an emerging population of high-energy radio galaxies}
Beyond giant radio galaxies, new radio surveys are unveiling a growing population of radio galaxies
associated with gamma-ray sources, in contrast with the common
picture that sees blazar as the almost unique counterpart.
Our group has recently studied a peculiar Fanaroff-Riley II INTEGRAL radio galaxy with a counterpart in the Fermi/LAT catalogue \citep{Bruni+2022}.
 Through a broad-band
spectral fitting from radio to gamma-ray, we found that the commonly invoked
jet contribution is not sufficient to account for the observed gamma-ray flux.
Instead, our modelling suggested that the observed emission could mainly
originate in the lobes (rather than in the radio core) by inverse Compton
scattering of radio-emitting electrons off the ambient photon fields. As
a follow-up of this work, we have proposed to join the MeerKAT+ survey with the goal of identifying Fermi
radio galaxy counterparts in the Southern hemisphere. Indeed, the survey will open a new window on the counterparts of the high-energy sky thanks to its unique combination of resolution and sensitivity.
The sound statistical basis provided by the survey will allow us to extend the SED
modelling of our recent pilot study to a complete sample of objects, with the goal of
unveiling the production site and process of gamma-ray emission.


\section{Conclusions}

The multi-scale/frequency study of radio galaxies, made possible by new generation radio surveys and telescopes, is revealing details about jets duty cycle at different accretion regime, and their evolution in the Mpc-scale environment. In particular, an emerging population of high-energy radio galaxies, detected by Fermi, is suggesting that ngVLA and SKA will be able to unveil more and more of these sources. VLBI observations can provides paramount insights on the central engine, revealing cases of dramatic jet reorientation and reactivation. The GRACE project will serve as a first step in exploring synergies between different instruments to probe jets formation and evolution in different systems.


\begin{acknowledgements}
GB acknowledges financial support for the GRACE project, selected via the Open Space Innovation Platform (\url{https://ideas.esa.int}) as a Co-Sponsored Research Agreement and carried out under the Discovery programme of, and funded by, the European Space Agency (agreement No.
4000142106/23/NL/MGu/my).
GB acknowledges financial support from the Bando Ricerca Fondamentale INAF 2023 for the project: \textit{\lq\lq The GRACE project: high-energy giant radio galaxies and their duty cycle\rq\rq}.
GB, AM, and MM acknowledge financial support from ASI under contract n. 2019-35-HH.0
This publication has received funding from the European Union’s Horizon 2020 research and innovation program under grant agreement No. 730562 (RadioNet). 
This publication has received funding from the European Union’s Horizon 2020 research and innovation programme under grant agreement No 101004719 (ORP). 
This work is partly based on observations with the 100-m telescope of the MPIfR in Effelsberg.
The European VLBI Network (www.evlbi.org) is a joint facility of independent European, African, Asian, and North American radio astronomy institutes. Scientific results from data presented in this publication are derived from the following EVN project code(s): {\tt{EB074}}.
e-MERLIN is a National Facility operated by the University of Manchester at Jodrell Bank Observatory on behalf of STFC. 
This work made use of the Swinburne University of Technology software correlator, developed as part of the Australian Major National Research Facilities Programme and operated under licence.
The Long Baseline Array is part of the Australia Telescope National Facility (https://ror.org/05qajvd42) which is funded by the Australian Government for operation as a National Facility managed by CSIRO.

\end{acknowledgements}


   \bibliographystyle{evn2024/aa} 
   \bibliography{evn2024/GRACE.bib} 


%
%
%
%
%
%
%
%
%
%
%
%

\end{document}